\documentclass{aa}  
\usepackage{graphicx}
\usepackage{epsfig}
\usepackage{lscape}
\usepackage{txfonts}
\begin{document}

\title{Analysis of possible anomalies in the QSO distribution 
of the Flesch \& Hardcastle catalogue}
\subtitle{}
\author{M. L\'opez-Corredoira\inst{1}, C. M. Guti\'errez\inst{1}, V.
Mohan\inst{2}, G. I. Gunthardt\inst{3,4}, M. S. Alonso\inst{5,6}}
\offprints{martinlc@iac.es}

\institute{
$^1$ Instituto de Astrof\'\i sica de Canarias, C/.V\'\i a L\'actea, s/n,
E-38200 La Laguna (S/C de Tenerife), Spain \\
$^2$ Inter-University Centre for Astronomy and Astrophysics (IUCAA), 
Post Bag No. 4, Ganeshkhind, Pune 411 007, India \\
$^3$ Observatorio Astron\'omico, Universidad Nacional de C\'ordoba, 
Laprida 854, 5000 C\'ordoba, Argentina \\
$^4$ SECyT, Universidad Nacional de C\'ordoba, Argentina \\
$^5$ Consejo Nacional de Investigaciones Cient\'\i ficas y T\'ecnicas, 
Argentina \\
$^6$ Complejo Astron\'omico El Leoncito, Argentina}

\date{Received xxxx; accepted xxxx}

  \abstract
   {}
   {A recent catalogue by Flesch \& Hardcastle presents two major anomalies
  in the spatial distribution of QSO candidates: $i/$ an apparent excess of
  such objects near bright galaxies, and $ii/$ an excess of very bright
  QSO candidates compared to random background expectations in 
  several regions of the sky. Because
  anyone of these anomalies would be relevant in a cosmological
  context, we carried out an extensive analysis of the probabilities
  quoted in that catalogue.}
   {We determine the nature and redshift of a subsample of 30 sources in
   that catalogue by analysing their optical spectra 
   (another 11 candidates were identified from existing public databases).
  These have allowed us to statistically check the reliability of the 
  probabilities QSO status quoted by Flesch \& Hardcastle for their candidates.}
   {Only 12 of the 41 candidates turned out QSOs
   (7 of which have been identified here for the first time).}
   {The probabilities of the QSOs' being the candidates
   given by Flesch \& Hardcastle are overestimated for $m_B\le 17$  and
   for objects projected near ($\le 1$ arcmin) bright galaxies. This is the
   cause of the anomalies mentioned above.}

\titlerunning{QSO candidates}
\authorrunning{L\'opez-Corredoira \& Guti\'errez}
   \maketitle
%

\section{Introduction}

At present, statistics on QSO distributions can be carried out
using the recent public compilations of QSOs identified in the SDSS
(Adelman-McCarthy et al.\ 2007) and 2dF (Croom et al.\ 2004) surveys. 
Furthermore, Flesch \& Hardcastle (2004; hereafter FH04) present an
all-sky catalogue with 86\,009 optical counterparts of radio/X-ray sources
as QSO candidates (with a
probability greater than 40\%, according to FH04) that were not
identified previously as such. This could be a useful catalogue to look 
for QSOs in regions that were not covered by SDSS or 2dF, or for
deeper magnitudes than their completeness limit. 
Nevertheless, since the FH04 catalogue does not
give the identification of each source but just a statistical estimate
of the probability that it be a QSO (or a star, 
a galaxy, or a wrongly identified optical
counterpart of the X-ray/radio source), it must be used with care.

Flesch\footnote{See his web-page
http://quasars.org/ .} claims that his catalogue `reveals a few
galaxy-centered fields which  are populated by QSOs/candidates more thickly
than expected  by the background density of such QSOs,' which would argue in
favour of the hypothesis of QSOs with anomalous redshift 
and at the same distance as the galaxies with which they are
associated (Arp 1998, Burbidge 2001).
Because this is against one of the most basic assumptions in standard modern
cosmology, we consider it interesting to check it on the basis of rigorous statistical
computation. This is the main objective of this paper. In 
\S \ref{.excesses}, we  analyse in which cases (on the basis of
the probabilities quoted by FH04)  there are anomalies in the distribution of
QSOs. By conducting optical spectroscopy (\S \ref{.select}), 
we directly check the nature of a subsample of FH04 sources.
On the basis of these observations, in  \S \ref{.probab} we
check the reliability of the probabilities quoted by FH04.
In \S \ref{.conclusions} we  reanalyse the distribution of QSOs and discuss
the implications of our results. 

\section{Anomalies in the distribution of QSO candidates}
\label{.excesses}

\subsection{Exploring the whole sky}
\label{.explosky}

The average number of QSOs up to $B$-magnitude $m_{B,0}$ 
is (L\'opez-Corredoira \& Guti\'errez 2006a, Eq. 2):

\[
N(m_B<m_{B,0})[{\rm deg}^{-2}]=
\]\begin{equation}
=\left \{ \begin{array}{ll}
        10^{-2.8+0.8(m_{B,0}-15)},& \mbox{ $m_{B,0}\le 18.5$} \\

        1981.0-214.2m_{b_j,0}+5.792m_{b_j,0}^2,& \mbox{ $m_{B,0}> 18.5$}
\end{array}
\right \}
\label{densQSO}
,\end{equation}
where the conversion of the photographic to the Johnson bandpass is
$m_{b_j,0}\approx m_{B,0}-0.14\times (m_{B,0}-m_{R,0})$.

Given a circular region of angular radius $R({\rm deg.})$ 
with $n$ QSO candidates up to magnitude $m_{B,0}$, 
the probability that FH04 expectations be compatible with
the expectations for background QSOs is

\begin{equation}
P=\sum _{m=0}^nP_p(m)P_q(m)
\label{P}
,\end{equation}
where $P_p(m)$ is the Poissonian probability
of  having $m$ sources in the circle

\begin{equation}
P_p(m)= \frac{[\pi R^2N(m_B<m_{B,0})]^m e^{-[\pi R^2N(m_B<m_{B,0})]}}{m!}
,\end{equation}
and $P_q(m)$ is the probability that we observe a 
number of $m$ QSOs among $n$ 
candidates given the a priori probabilities $p_1,...,p_n$:

\begin{equation}
P_q(m<n)=P_q(n)\sum _{l_1,...,l_{n-m}=1; l_1<...<l_{n-m}}^n
[(p_{l_1}^{-1}-1)...(p_{l_{n-m}}^{-1}-1)]
\label{probpriori}
,\end{equation}\[
P_q(n)=\prod _{i=1}^n p_i
.\]
The probability of finding at least one region with low probability
$P$ in the whole sky is
\begin{equation}
P_{\rm sky}=P\times (41252\ {\rm deg}^2)\times {\rm Minimum}
\left[\frac{1}{\pi R^2},N(m_B<m_{B,0})\right]
.\end{equation}

We have explored the distribution of QSO candidates in several projected
circles on the sky to check their compatibility with the expectation from
background QSOs. 
In Table \ref{Tab:excessmag}, we give the centre and radius of 
the regions and magnitudes that are very unlikely ($P_{\rm sky}<0.05$, 
i.e. 95\% C.L.) as a result of  chance projection.  
The table shows many of these cases.\footnote{These probabilities are
conservative because: i/ the catalogue ignores QSO candidates with
probabilities $\le 40$ \%; ii/ it includes only those QSO candidates 
that were not classified previously; iii/ it only includes
QSOs that emit in radio or X-ray, which is not the whole QSO population;
iv/ it is not complete for the 
faintest fluxes because the radio and X-ray catalogues used do not have
the same depth all over the sky.}

\begin{table}
\caption{Probabilities of QSO excesses.}
\begin{center}
\begin{tabular}{cccccc}
\label{Tab:excessmag}
$\alpha (^\circ )$, $\delta (^\circ )$ 
& R ($^\circ $) & $m_{B,0}$ &  $n$ & $P$ & $P_{\rm sky}$ \\ \hline
324.534,  -66.000 &  6 & 13.0  &  2 & 0.026 & 0.042 \\
324.534,  -66.000 &  6 & 13.5  &  3 & 0.010 & 0.041 \\
 30.193,  -58.000 & 16 & 14.0  &  4 & 0.0041 & 0.042 \\
 58.476,  -70.000 & 20 & 13.0  &  2 & 0.021 & 0.035 \\
 59.006,  -66.000 & 24 & 13.0  &  2 & 0.023 & 0.038 \\
 62.229,  -50.000 & 40 & 13.0  &  3 & 0.0 21 & 0.035 \\
 62.229,  -50.000 & 40 & 14.0  &  6 & 0.0034 & 0.036 \\
 63.640,  -45.000 & 45 & 13.0  &  3 & 0.023 & 0.038 \\
 63.640,  -45.000 & 45 & 14.0  &  7 & 0.0021 & 0.022 \\
 63.640,  -45.000 & 45 & 14.5  &  12 & 0.0013 & 0.027 \\
 65.270,  -40.000 & 50 & 13.0  &  3 & 0.026 & 0.043 \\
 65.270,  -40.000 & 50 & 13.5  &  5 & 0.0071 & 0.029 \\
 65.270,  -40.000 & 50 & 14.0  &  8 & 0.0014 & 0.014 \\
 65.270,  -40.000 & 50 & 14.5  &  16 & $3.1\times 10^{-4}$ & 0.0052 \\
 67.143,  -35.000 & 55 & 14.0  &  8 & 0.0041 & 0.043 \\
 67.143,  -35.000 & 55 & 14.5  &  16 & 0.0011 & 0.015 \\
 69.282,  -30.000 & 60 & 14.5  &  20 & $2.3\times 10^{-4}$ & 0.0027 \\
 71.720,  -25.000 & 65 & 14.5  &  19 & $8.6\times 10^{-4}$ & 0.0083 \\
 69.282,  -30.000 & 60 & 12.5  &  3 & 0.058 & 0.038 \\
 71.720,  -25.000 & 65 & 12.5  &  3 & 0.060 & 0.039 \\
 65.270,  -40.000 & 50 & 15.0  &  20 & 0.0020 & 0.033 \\
 67.143,  -35.000 & 55 & 15.0  &  22 & 0.0028 & 0.038 \\
 69.282,  -30.000 & 60 & 15.0  &  27 & $9.7\times 10^{-4}$ & 0.011 \\
 71.720,  -25.000 & 65 & 15.0  &  27 & 0.0036 & 0.035 \\
 84.783,  -64.200 &0.3 & 19.0  &  20 & $6.3\times 10^{-8}$ & 0.0055 \\
 84.860,  -64.000 &0.4 & 19.0  &  26 & $6.2\times 10^{-9}$ & $5.4\times 10^{-4}$\\
 84.860,  -64.000 &0.4 & 19.5  &  30 & $4.7\times 10^{-8}$ & 0.012 \\
 84.403,  -64.000 &0.5 & 19.0  &  25 & $3.1\times 10^{-8}$ & 0.0027 \\
 84.936,  -64.100 &0.7 & 19.0  &  29 & $4.6\times 10^{-8}$ & 0.0039 \\
 85.168,  -64.400 &0.8 & 19.0  &  30 & $7.0\times 10^{-7}$ & 0.0045 \\
\end{tabular}
\end{center}
{\bf Note:} $P$ is the probability of there being $n$ QSO candidates up to
magnitude $m_{B,0}$ in the FH04 catalogue
with the probabilities specified by them within a circle of centre
$(\alpha , \delta )$ (coordinates J2000) and angular radius $R$. 
$P_{\rm sky}$ is the probability
of finding at least one circle with the small value of $P$ in the whole sky.
Cases with $P_{\rm sky}<0.05$.
\end{table}

The large excess of QSO candidates up to B-magnitude $\sim 19$ in the
region centred at  $\alpha \approx 84.9^\circ =5h39m36s$, $\delta =-64^\circ $ and radius
0.4 deg (26 QSO candidates in a 0.50 deg$^2$ region up to magnitude 19
while an average of $\sim 1$ QSO is expected as background)
might be related with the Large Magellanic Clouds (LMC),
whose centre is 7 degrees away. 
The region around LMC contains a number of X-ray
sources (Pietsch \& Kahabka 1993; Haberl \& Pietsch 1999), many 
of them background QSOs (Kahabka 2002; Dobrzycki et al. 2005), but
many of them are other kinds of objects that could have been confused with
QSO candidates by FH04.
Particularly, there is a very bright X-ray source exactly in this region: 
LMC X-3 (Haberl \& Pietsch 1999) with 
$\alpha =5h38m56.8s$, $\delta =-64^\circ 05'12" $, a black hole
binary, a Be star with a period 1.7 days. Most probably,
the many ROSAT X-ray sources surrounding it might be
attenuated reflections (by the X-ray detector, which might create
ghost images when a strong source saturates it)
of this very bright X-ray source.

However, the excess of very bright QSOs in many areas given in
Table \ref{Tab:excessmag} cannot be explained by any similar contamination.
For instance, in the circle centred at
$\alpha \approx 69.282^\circ =4h37m7.7s$, $\delta =-30^\circ $ and radius
60 deg. [Twenty QSO candidates in a 1/4 of the whole sky up to magnitude
14.5, while an average of $\sim 7$ QSOs is expected as background].

\subsection{Exploring the regions near bright galaxies}
\label{.explorc3}

\begin{figure}
\vspace{1cm}
{\par\centering \resizebox*{6cm}{6cm}{\includegraphics{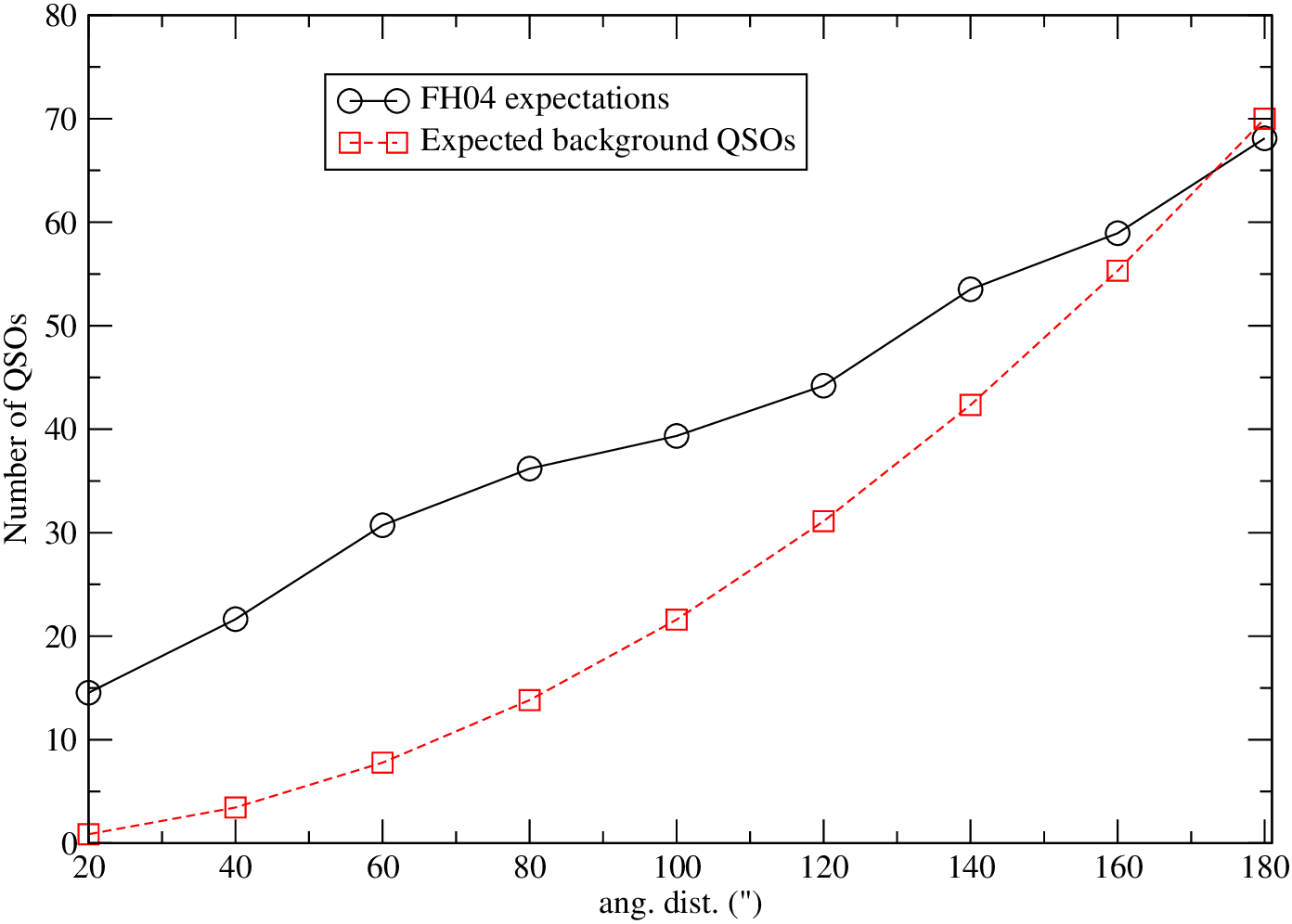}}\par
\vspace{1.2cm}
\resizebox*{6cm}{6cm}{\includegraphics{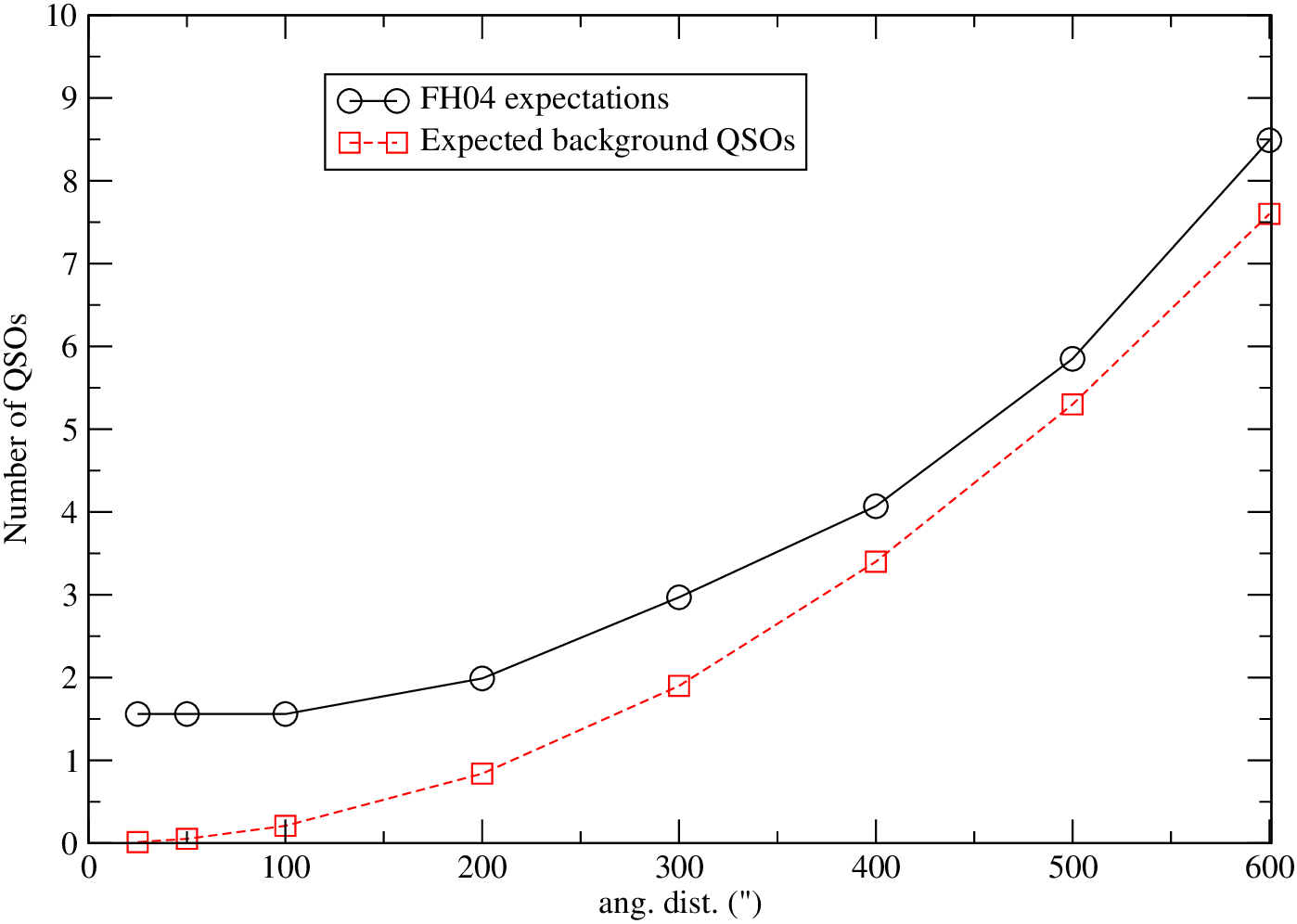}}\par}
\caption{Cumulative counts of FH4 QSOs (sum of the probabilities 
given by FH04 to be QSOs) around the 
complete sample of RC3 galaxies as a function of the maximum 
distance to a galaxy, in comparison with the expected background QSOs. 
Up: up to magnitude $m_B=18$, Down: up to magnitude $m_B=15.5$.
QSOs associated to two galaxies count twice
with each corresponding distance.}
\label{Fig:QSOdist}
\end{figure}

Here we explore the regions near bright RC3 galaxies (Third Reference
Catalog of Bright Galaxies; de Vaucouleurs et al. 1991, updated on 1995,
February 16th). The RC3 catalogue is complete for galaxies with $m_{B,gal}<15.5$,
redshift $z\le 0.05$, and an apparent diameter at the $D_{25}$ isophotal level
larger than  1 arcminute. The catalogue contains a total of 23,011 galaxies.
From Eq. (\ref{densQSO}), around all the RC3 galaxies and
within a distance of $d$ arcminutes and magnitude $m_B<m_{B,0}$, 
we would expect a number of QSOs:

\begin{equation}
N_{\rm RC3}(m_{B,0},d('))\approx 10^{0.8(m_{B,0}-16.9)}d(')2
\label{densrc3}
,\end{equation}
which means an average of 70 objects up to 
3 arcminutes of RC3-galaxies with $m_B\le 18$ and 
7.6 cases up to 10 arcminutes with $m_B\le 15.5$. 
Figure \ref{Fig:QSOdist} shows how FH04 average expectations
exceed the expected background counts,
and Table \ref{Tab:QSOdist} shows the very low probability
[calculated with Eqs. (\ref{P}) and (\ref{densrc3})] of
making both numbers compatible, specially for $d<1'$ and 
$m_B\le 18$. 
For the highest $d$ values, the probability is somewhat lower due to the
incompleteness of the FH04 catalogue rather than to an excess in the number of
QSOs.

\begin{table}
\caption{Probabilities of QSO excesses around RC3 galaxies.}
\begin{center}
\begin{tabular}{ccccc}
\label{Tab:QSOdist}
$d(")$ & $m_{B,0}$ & $n$ & $\langle n_{bg}\rangle $ & $P_{\rm RC3}$  \\ \hline
0-20 & 18.0  &   25 & 0.86 & $3.5\times 10^{-7}$  \\
20-40 & 18.0  &   12 & 2.6 & 0.027  \\
40-60 & 18.0  &   14 & 4.3 & 0.031  \\
60-80 & 18.0  &   8 & 6.0 & 0.15  \\
80-100 & 18.0  &  5 & 7.8 & 0.042  \\
100-120 & 18.0  &  8 & 9.5 & 0.049  \\
120-140 & 18.0  &  13 & 11 & 0.10  \\
140-160 & 18.0  &  8 & 13 & 0.013  \\
160-180 & 18.0  &  13 & 15 & 0.042  \\
0-25 & 15.5  &   3 & 0.013 & 0.096  \\
0-50 & 15.5  &   3 & 0.053 & 0.11  \\
0-100 & 15.5  &   3 & 0.21 & 0.14  \\
100-200 & 15.5  &   1 & 0.63 & 0.45  \\
200-300 & 15.5  &   2 & 1.1 & 0.32  \\
300-400 & 15.5  &   2 & 1.5 & 0.29  \\
400-500 & 15.5  &  3 & 1.9 & 0.25  \\
500-600 & 15.5  &  3 & 2.3 & 0.22  \\
\end{tabular}
\end{center}
{\bf Note:} $P_{RC3}$ is the probability of having $n$ 
QSO candidates up to magnitude $m_{B,0}$ in FH04 catalogue
with the probabilities specified by them within a circle centred
at a RC3 galaxy and angular distance $d$.
\end{table}

\begin{figure*}
\vspace{1cm}
{\par\centering \resizebox*{12cm}{17.3cm}{\includegraphics{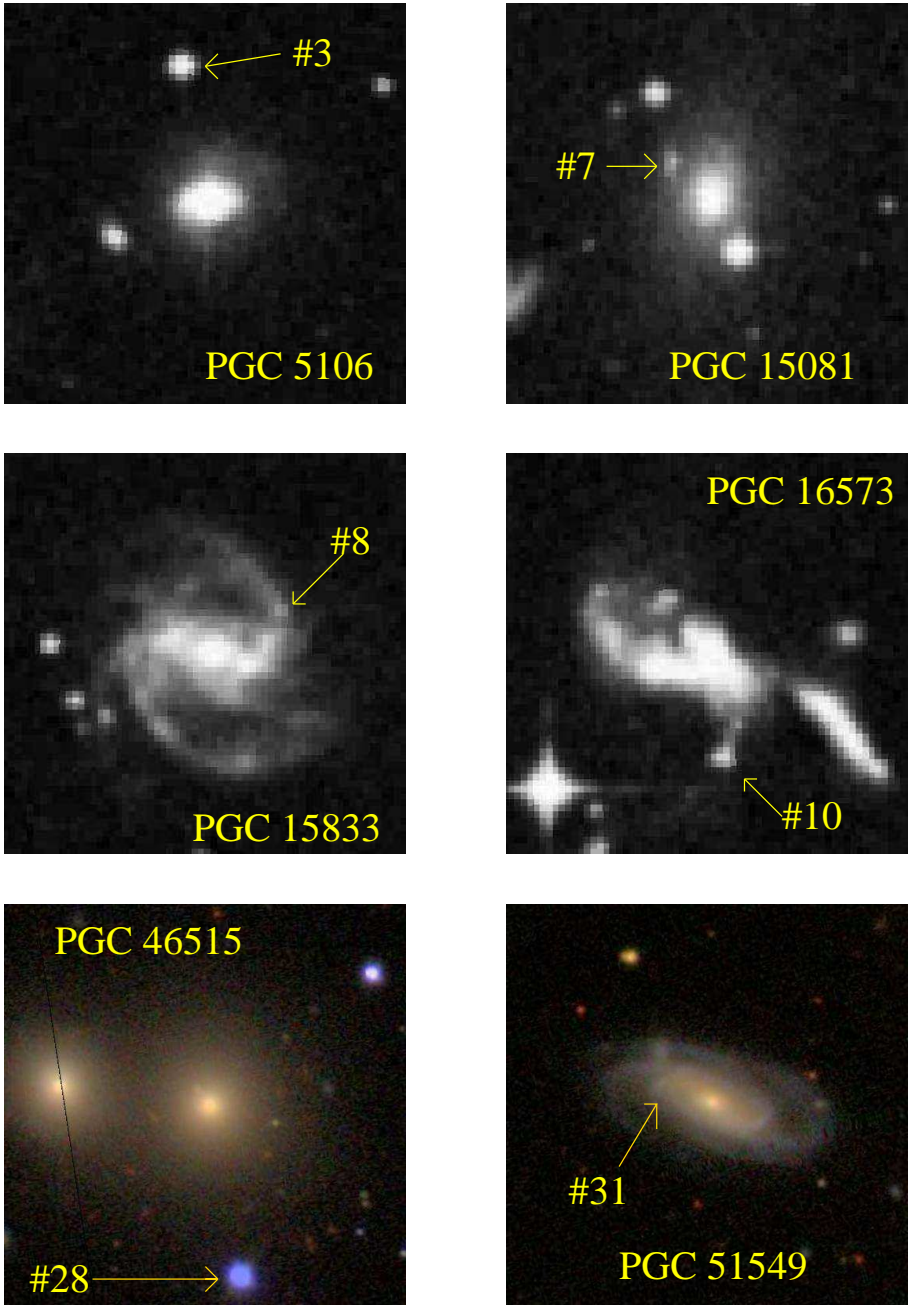}}}
\caption{Some of the QSO candidates of FH04 observed in the present
work (number \# as in Table \protect{\ref{Tab:objects}}). 
PGC 5106, PGC 15081, PGC 15833, PGC 16573 images 
from the Digitized Sky Survey, size $108"\times 108"$. 
PGC 46515, PGC 51549 images 
from the Sloan Digital Sky Survey, size $100"\times 100"$. 
None 
of these candidates is a QSO.}
\label{Fig:examples}
\end{figure*}

Among the QSO candidates that are less than 1 arcminute from a 
RC3 galaxy, we have the examples shown in Figure \ref{Fig:examples}.

\section{Selection and observation of QSO candidates}
\label{.select}

The anomalies in \S \ref{.excesses} rely on the probabilities quoted by FH04.
To check the reliability of these probabilities, we determined 
the nature and redshifts of a significant sample of FH04 objects
by direct observations .
The objects selected follow four criteria:

\begin{itemize}

\item a) Sources up to 3 arcminutes from RC3 galaxies with $m_B\le 18$, such
as those in Figure \ref{Fig:examples}.

\item b) Sources up to 10 arcminutes from RC3 galaxies 
with $m_B\le 15.5$. 
 
\item c) Other sources up to 3 arcminutes from galaxies with $m_B\le 18$
that do not belong to the RC3 catalogue.

\item d) Sources with $m_B\le 15.5$ (to test the anomalous distribution
of this bright sources in several regions of the sky, as shown in Table
\ref{Tab:excessmag}).

\end{itemize}

In total, we have identified the 41 sources listed in Table
\ref{Tab:objects}: 11 from the public database of SIMBAD (data obtained
after the publication of FH04) and 30 from
our own spectroscopic observations. A double check was made by observing 
5 objects already classified in SIMBAD. The columns of Table
\ref{Tab:objects} are as follows.
Column 1 gives number of the source. Column 2: Equatorial 
coordinates (J2000) of the optical counterpart
identified by FH04 as a QSO candidate. Column 3: apparent magnitude in the
$B$ filter (FH04). Column 4: probability that the source is a QSO according to FH04.
Column 5: Radio flux at 1.4 GHz (mJy) or flux in X-ray 
(counts/hour in ROSAT) [FH04].
Column 6: peculiarity of the source: a) sources up to 3 arcminutes from RC3 galaxies with $m_B\le 18$; 
b) sources up to 10 arcminutes from RC3 galaxies with $m_B\le 15.5$; c) sources
up to 3 arcminutes from non RC3 galaxies with $m_B\le 18$; d) sources with $m_B\le 15.5$. 
In cases a), b), and c): the RC3 galaxy (or non-RC3 for case \# 6)
from which the source is closest, and the distance (in arcseconds)
between the QSO candidate and the galaxy are specified.
Column 7: Exposure time in seconds (with the superindex $^{np}$ means 
non-photometric),
date of observation and telescope (CA: 2.15 m CASLEO, San Juan, Argentina;
OH: 1.93 m OHP, Haute-Provence, France; IU: 2 m IUCAA, India) or
``SIMBAD-database'' when the data were publicly available on the SIMBAD
web-page. Column 8: Spectroscopic identification. Column 9: Spectral
lines:  CIV is at 1549 \AA , CIII is at 1909 \AA , MgII is at 2798 \AA ,
NeIII is at 3869 \AA , HeI is at 3889 \AA , CaII is the doublet
H and K at 3933/3970 \AA , H$_\delta $ is at
4102 \AA , H$_\gamma $ is at 4340 \AA , HeII is at 4686 \AA , H$_\beta $ is at 4861 \AA , 
OIIIa is at 4959 \AA , OIIIb is at 5007 \AA , MgI is at 5180 \AA , NaI is at 5892 \AA ,
OI is at 6300 \AA , NIIa is at 6548 \AA , H$_\alpha $ is at 6562 \AA , NIIb is at 6584 \AA,
SIIa is at 6717 \AA , SIIb is at 6731 \AA ; ``A:'' means absorption, ``E:'' means 
emission, superindex $b$ means ``broad''. Column 10: redshift.

We used the following telescopes and
instrumental setups:

\begin{itemize}

\item 2.15 m CASLEO, in ``El Leoncito'' Observatory,
San Juan, Argentina. REOSC Ds Spectrograph, 
grism \# 270, slit width: 2.5$"$. Wavelength range: 3630--7120 \AA ,
3.41 \AA /pixel.

\item 1.93 m OHP, Haute-Provence, France. 
CARELEC Spectrograph, whose grating  provides a resolution of 
133 \AA /mm, slit width: 2$"$.
1.78 \AA /pixel. Wavelength range: 3965--7600 \AA . 

\item 2 m IUCAA, India. 
2 m f/10 Cassegrain telescope. IFOSC (IUCAA Faint Object
Spectrograph and Camera), f/4.5. Grism FOSC7, wavelength range
3800--6840 \AA , dispersion 1.39 \AA /pixel, slit width: 1.5$"$.

\end{itemize}

\begin{figure}
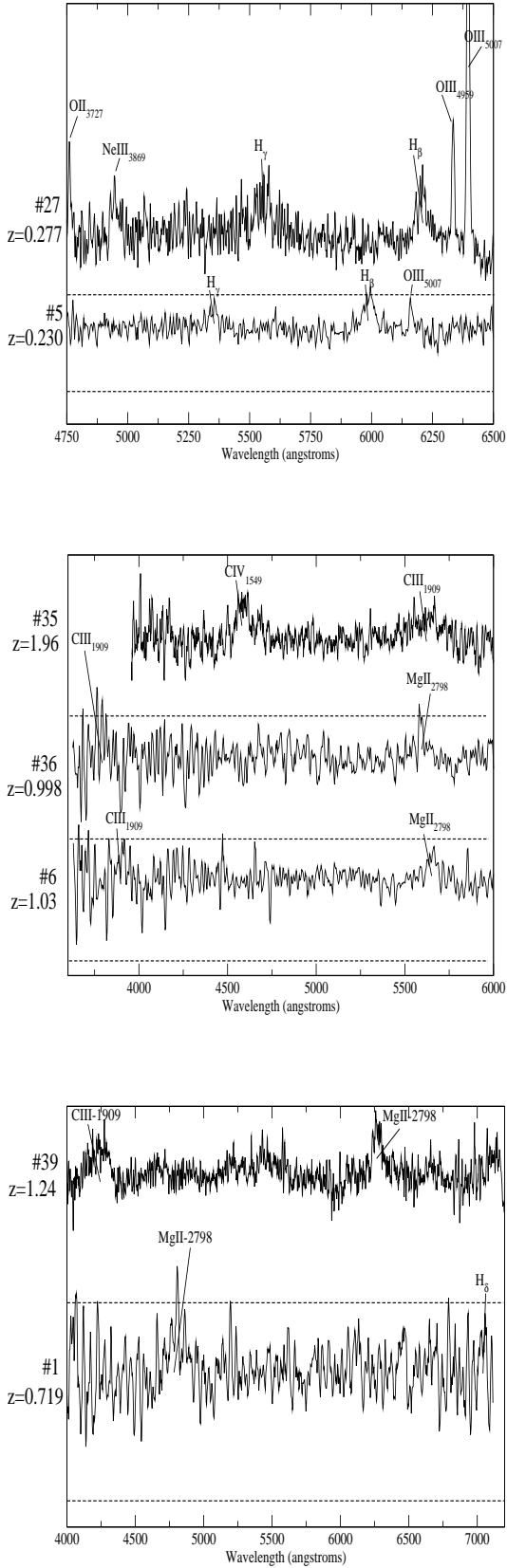

\vspace{1cm}
{\par\centering \resizebox*{7cm}{6.5cm}{\includegraphics{spectra1.eps}}\par
\vspace{1.2cm}
\resizebox*{7cm}{6.5cm}{\includegraphics{spectra2.eps}}\par
\vspace{1.2cm}
\resizebox*{7cm}{6.5cm}{\includegraphics{spectra3.eps}}\par}
\caption{Continuum normalized spectra of the seven objects 
classified as QSOs from a total of 30 FH04 candidates. 
Dashed lines stand for zero flux.}
\label{Fig:spectra}
\end{figure}

We followed a standard method of reduction. A summary of the identifications is
presented in Table \ref{Tab:objects}. Only 7 of the 30 objects turned out to be  QSOs,
while the other 23 were stars, HII regions, or galaxies. The spectra of the seven QSOs are plotted in Figure
\ref{Fig:spectra}. For the 11 SIMBAD database objects, there were
another 5 QSOs, so we found 12 QSOs from a total number of 41 objects observed.

\subsection{Notes on some objects}

\begin{description}

\item [\#7]: this is 
a Seyfert 2 galaxy at only 13 kpc projected distance from
a cD galaxy (PGC 15081) within the same cluster.

\item [\#8,\#24, \#31]: these are star-forming regions that were probably 
considered as QSO candidates on the basis of their blue colours.

\item [\#11,\#16, \#20, \#32]: white dwarfs are also usually
misidentified as QSO candidates because of their blue
colours. 

\item [\#15]: this is a cold M5 star with emission lines
presumably caused by photospheric activity, which gives strong
X-ray emission.

\end{description}

The rest of the non-QSO objects are other types of stars. The number
of contaminating objects in case a) [up to magnitude
18 within 180 deg$^2$] or b) [up to magnitude 15.5 within 2000
deg$^2$] is not surprising since the number of stars in
high galactic latitude fields is $\sim 10^5-10^6$ (Robin et
al. 2003) for both cases a) and b), and among them a fraction
$\sim 10^{-4}-10^{-5}$ might be white dwarfs or stars with some activity
or feature that make them susceptible to being detected in
radio or X-ray. 

\section{The reliability of FH04 probabilities}
\label{.probab}

\subsection{As a function of the apparent magnitude}

\begin{figure}
\vspace{1cm}
\resizebox*{7cm}{7cm}{\includegraphics{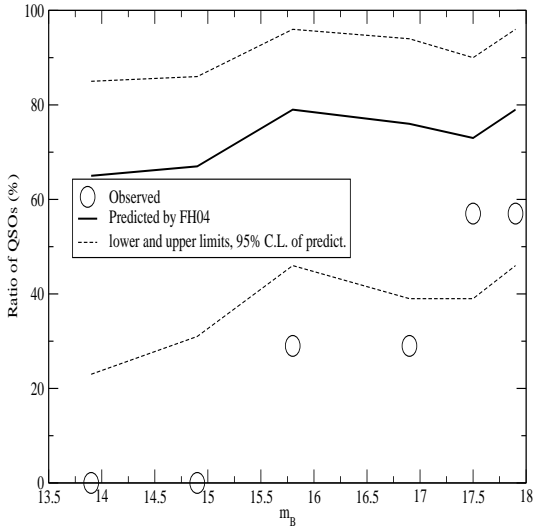}}\par
\caption{Observed ratio of QSOs as a function of magnitude among
the 41 FH04 candidates with spectroscopic identification given
in this paper. Comparison with predictions of FH04 (limits within
dashed line are 95\% C.L.).}
\label{Fig:prob}
\end{figure}

Given the observed number of QSOs that have different magnitudes
with the a priori probabilities of FH04 (4th column of 
Table \ref{Tab:objects}),
we can calculate $P(\le m)$ with Eq. (\ref{probpriori}), the
probability that FH04 expectations are correct given that
we have detected a number of $\le m$ QSOs among the $n$ candidates.
They are given in Table \ref{Tab:probfleschcorrect}.
Clearly,  for magnitudes $m_B<\approx 17$, the a priori
probabilities given by FH04 are overestimated.
However, for $m_B>\approx 17$ our results
are compatible with the probabilities given by FH04.
Figure \ref{Fig:prob} and Table \ref{Tab:probfleschcorrect} show
the expected and the observed average probabilities of the FH04 candidates
being QSOs. 

Given that none of the 13 QSO candidates
with $m_B\le 15.1$ are in fact QSOs, this means that the excesses 
of Table \ref{Tab:excessmag} with $m_B\le 15$ are not real QSO excesses,
so there is no incompatibility with background expectations.

\begin{table}
\caption{Probabilities that FH04 expectations are correct as a function
of the magnitude.}
\begin{center}
\begin{tabular}{ccccc}
\label{Tab:probfleschcorrect}
$m_B$ & $n$ & $m$ & 
$m_{\rm FH04}$  & $P(\le m)$ 
\\ \hline

13.4-14.4 & 6 & 0 & 3.9$^{+1.2}_{-2.5}$ & $1.1\times 10^{-4}$ \\

14.7-15.1 & 7 & 0 & 4.7$^{+1.3}_{-2.5}$ & $5.2\times 10^{-5}$ \\

15.2-16.5 & 7 & 2 & 5.5$^{+1.2}_{-2.3}$ & $2.4\times 10^{-3}$ \\

16.7-17.2 & 7 & 2 & 5.3$^{+1.3}_{-2.6}$ & 0.011 \\

17.3-17.7 & 7 & 4 & 5.1$^{+1.2}_{-2.4}$ & 0.23 \\

17.8-18.0 & 7 & 4 & 5.5$^{+1.2}_{-2.3}$ & 0.14 \\

\end{tabular}
\end{center}
{\bf Note:} Column 2: $n$: number of
FH04 QSO candidates observed. Column 3: $m$ number of these which are
identified spectroscopically as QSOs. Columm 4: expected number
of QSOs according to FH04 ($\pm $; 95\% C.L.). Column 5:
$P(\le m)$, probability that FH04 expectations are correct given
that we have detected a number $\le m$ of QSOs among the $n$ candidates.
\end{table}

\subsection{As a function of the distance to an RC3 galaxy}

\begin{figure}
\vspace{1cm}
\resizebox*{7cm}{7cm}{\includegraphics{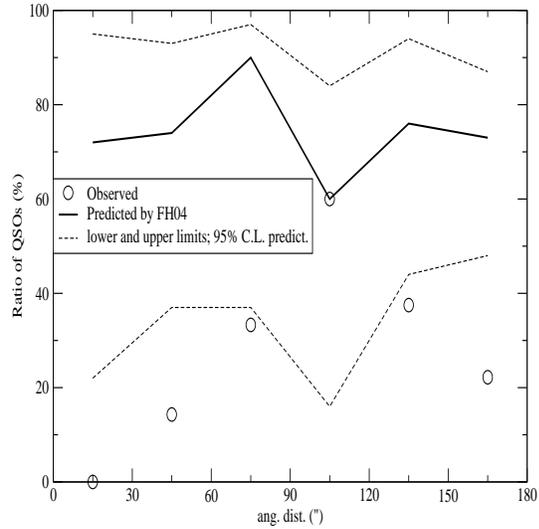}}\par
\caption{Observed ratio of QSOs as a function of distance to an
RC3 galaxy among the 36 FH04 candidates (including repetitions, i.e. objects
associated with more than one galaxy) 
with spectroscopic identification given
in this paper. Comparison with the predictions of FH04 (limits within
dashed line are 95\% C.L.).}
\label{Fig:prob2}
\end{figure}

For the subsample a) of Table \ref{Tab:objects}, there are
26 objects with $m_B\le 18$ at distances less than
180$"$ from an RC3 galaxy, 36 cases if we count the repetitions.\footnote{There 
are some cases of two RC3 galaxies projected 
very close in the sky. For these cases QSOs projected near such 
galaxies are counted twice (we will refer to
these cases as `cases with repetition'), so this raises
the number of QSO/galaxy associations to 36.}
Table \ref{Tab:probfleschcorrect2} gives the probability $P(\le m)$
that FH04 expectations are correct given that
we detected a number of $\le m$ QSOs among the $n$ candidates,
as a function of the distance of the RC3 galaxy.
Figure \ref{Fig:prob2} and Table \ref{Tab:probfleschcorrect2} show
the expected and the observed average probability of the FH04 candidates
being QSOs. We conclude that the FH04 probabilities 
are highly overestimated in
the regions within $\sim 1$ arcminute from an RC3 galaxy. There
is also overestimation at distances $\sim 2$--3 arcminutes.

This result solves the second anomaly.
Since only 1 object among 11 candidates at distances less than 1 
arcminute is really a QSO, the probabilities given by FH04 are not
valid, and the announced excess announced 
in \S \ref{.explorc3} 
is not supported. 

\begin{table}
\caption{Probabilities that FH04 expectations are correct as a function
of the distance to RC3 galaxies.}
\begin{center}
\begin{tabular}{ccccc}
\label{Tab:probfleschcorrect2}
d(") & $n$ & $m$ & $m_{\rm FH04}$ & $P(\le m)$ 
\\ \hline

0-30 & 4 & 0 & 2.9$^{+0.9}_{-2.0}$ & $4.2\times 10^{-3}$ \\

30-60 & 7 & 1 & 5.2$^{+1.3}_{-2.6}$ & $1.1\times 10^{-3}$ \\

60-90 & 3 & 1 & 2.7$^{+0.2}_{-1.6}$ & 0.022 \\

90-120 & 5 & 3 & 3.0$^{+1.2}_{-2.2}$ & 0.68 \\

120-150 & 8 & 3 & 6.1$^{+1.4}_{-2.6}$ & 0.015 \\

150-180 & 9 & 2 & 6.6$^{+1.2}_{-2.3}$ & $8.8\times 10^{-4}$ \\

\end{tabular}
\end{center}
{\bf Note:} Column 1: range of distances from an RC3 galaxy. Column 2: $n$: number of
FH04 QSO candidates observed. Column 3: $m$ number of these which are
identified spectroscopically as QSOs. Columm 4: expected number
of QSOs according to FH04 ($\pm $; 95\% C.L.). Column 5:
$P(\le m)$, probability that FH04 expectations are correct given
that we have detected a number $\le m$ of QSOs among the $n$ candidates.
\end{table}

\section{Discussion and conclusions}
\label{.conclusions}

In FH04 there are 88 different objects (106 cases
with repetitions)
with $m_B\le 18$ within a radius of 3 arcminutes of RC3 galaxies.
Of these 88 objects, there are spectroscopic observations
of the optical counterparts in 26 cases (36 cases including repetitions),
and 10 of these (the same number with repetitions) are
in fact QSOs
(while the 16 other objects (26 with repetitions) are stars, galaxies,
or HII regions).
The average expected number of background cases around all RC3
cases should be $\approx 70$ (\S \ref{.explorc3}) and, in our subsample,
should be $\approx \frac{36}{106}\times 70=24$ cases, which is higher than
the value derived from the observations. 
This stems from the incompleteness of the FH04 catalogue.

Within a radius of 10 arcminutes of RC3-galaxies, there are
14 objects (no repetitions) with $m_B\le 15.5$, of which five
(36\% of the sample) were
observed and only one (\# 5 with $m_B=15.4$) turned out a QSO.
The average expected number of
cases around all RC3 galaxies is 7.6 (\S \ref{.explorc3}), and
in our subsample should be $\frac{5}{14}\times 7.6=2.7$ cases.
This result agrees with expectations from a background distribution 
of QSOs.

Limiting the study to RC3 galaxies by $m_{B,gal}\le 14.5$,
$m_{B,gal}\le 13.5$, and $m_{B,gal}\le 12.5$, the number of galaxies 
is reduced to 7,675/2,566/883.
For the subsample with $m_b\le 18$, $d\le 3'$ we obtained
5/2/2 QSOs to be compared with the expected 7.8/2.7/0.92.
For the subsample with $m_b\le 15.5$, $d\le 10'$, we obtained
1/0/0 to be compared with 0.50/0.17/0.06.
There is no statistical
inconsistency with the claim that all QSOs are background QSOs rather
than  being associated with the galaxies.

\begin{figure}
\vspace{1cm}
\resizebox*{7cm}{7cm}{\includegraphics{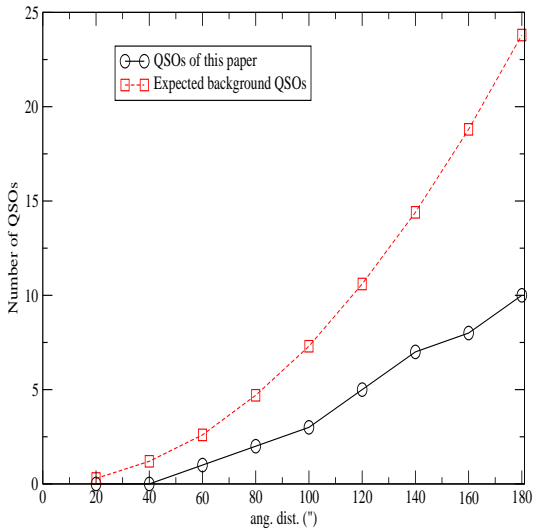}}\par
\caption{Cumulative counts of QSOs of this paper
up to magnitude $m_B=18$ [34\% (36/106) of the 
QSO/RC3--galaxy pairs] as a function of the maximum 
distance to the galaxy. QSOs associated with two galaxies count twice
with each corresponding distance.}
\label{Fig:QSOdist2}
\end{figure}

As for the distribution of the QSOs with magnitude or distance from RC3
galaxies, as pointed out in \S \ref{.probab}, there is no longer 
any inconsistency with background predictions: 
almost all the confirmed QSOs in our subsample are faint
(all of them with magnitudes $m_B\ge 17.0$, except \#5)  and distant enough from
the centres of RC3 galaxies to be compatible with background expectations. See
Fig. \ref{Fig:QSOdist2} for the distance distribution and note how the observed
QSOs are fewer than the background expectations for all distances.
Some of the most
troublesome candidates, such as those shown in Fig. \ref{Fig:examples}, were in the
end confirmed to be objects different from QSOs.

The only QSO with $m_B\le 15.5$ (\# 5)
within 10$'$ from an RC3 galaxy is associated with
NGC 1136, which was previously reported as a galaxy
with peculiar associations (Arp 1981). Objects claimed by Arp to be anomalous
are much fewer than 23,011 RC3 galaxies. They are $\sim 10^2$,
so the probability of finding a QSO with $m_B\le 15.5$
within 10$'$ from some of them is $P\sim 0.06$. It is not few enough to claim
it separately from other facts as a possible anomalous case,
but a fact to be added in the discussion
of the case of NGC 1136.

For the object \# 6, now confirmed to be a QSO with $m_B=16.5$
at only 30$"$ from PHL 1459 (a galaxy with $m_B=17.2$ and X-ray emission),
we are not within the statistics of RC3 galaxies.
There are $\approx 4.3\times 10^5$ galaxies in the whole sky up
to the magnitude $m_B=17.2$ (Metcalfe et al.\ 1991), so, using
Eq.\ (\ref{densQSO}), the expected number of cases
like this within a 30-arcsecond distance of these galaxies is 2.2. The discovery of
only one case is not enough to claim any statistical
anomaly. If we take into account the peculiarity that PHL 1459 is
an X-ray source and has $\approx 120$ counts/hour in ROSAT (Voges
et al.\ 1999, 2000), the probability is somewhat lower.
There are $\sim 0.4$ galaxies/deg$^2$ up to this
X-ray flux (Voges et al. 1999; Zickgraf et al. 2003), i.e. around 16,000 galaxies
like this in the whole sky. The probability of finding a QSO
with $m_B=16.5$ within 30$"$ of one of them is 0.09;
not low enough to discard a background projection coincidence.

Summing up, within the subsample of 41 objects from FH04, we have not found 
any statistical anomaly with respect to the expectation of QSOs as background
sources. Since we observed a representative subsample of
the most controversial cases that could in principle present some excesses of
QSOs, we think that there should be in principle no reason 
to expect statistical anomalies in the FH04 catalogue, which 
is anyway a useful database for looking for QSOs in regions not
covered by other surveys, such as SDSS or 2dF, or for deeper
magnitudes. Nevertheless, there are still anomalies in the QSO distributions
presented with some other catalogues (e.g.\ Arp 1998; Burbidge 2001) that require
further attention. Further rigorous statistical analysis like those presented
here and in L\'opez-Corredoira \& Guti\'errez (2006$a,b$) are necessary.

\

{\bf Acknowledgments:} 

Thanks are given to Rub\'en J. D\'\i az (Gemini Observatory)
for providing telescope time
at 2.15 m CASLEO, to Eric Flesch for his comments and criticism 
on a draft of this paper, to the referee Mira Veron for her
helpful comments for improving it,
to T. J. Mahoney (IAC) and Joli Adams (A\&A) 
for proofreading the paper.

Based on observations made with the telescopes: 2.15 m CASLEO in the
observatory of ``El Leoncito'', San Juan, Argentina;
1.93 m OHP (these observations were funded by the Optical Infrared 
Coordination network, OPTICON, a major international 
collaboration supported by the Research 
Infrastructures Programme of the European Commission Sixth Framework 
Programme), Haute-Provence, France; and 2 m IUCAA (Pune, India).

This research has made use of the SIMBAD database,
operated at CDS, Strasbourg, France.

The Digitized Sky Surveys were produced at the Space Telescope Science 
Institute under US Government grant NAG W-2166. The images of these 
surveys are based on photographic data obtained using the Oschin Schmidt 
Telescope on Palomar Mountain and the UK Schmidt Telescope. 
The plates were processed into the present compressed digital form with 
the permission of these institutions.

Funding for the creation and distribution of the SDSS Archive has  been
provided by the Alfred P. Sloan Foundation, the Participating  Institutions, the
National Aeronautics and Space Administration, the  National Science Foundation,
the U.S. Department of Energy, the Japanese  Monbukagakusho, and the Max Planck
Society. The SDSS Web site is  http://www.sdss.org/. The SDSS is managed by the
Astrophysical Research  Consortium (ARC) for the Participating Institutions. The
Participating  Institutions are The University of Chicago, Fermilab, the
Institute for  Advanced Study, the Japan Participation Group, The Johns Hopkins
University,  the Korean Scientist Group, Los Alamos National Laboratory,  the
Max-Planck-Institute for Astronomy (MPIA), the Max-Planck-Institute for 
Astrophysics (MPA), New Mexico State University, University of Pittsburgh, 
University of Portsmouth, Princeton University, the United States Naval 
Observatory, and the University of Washington.

The first author (MLC) was supported by the {\it Ram\'on y
Cajal} Programme of the Spanish Science Ministry.

\eject

\begin{landscape}

\begin{table}
\caption{Observed FH04 QSO candidates (column description in \S 
\ref{.select}).}
\begin{center}
\begin{tabular}{cccccccccc}
\label{Tab:objects}
\# & $\alpha (h, m, s),\delta (d, ', ")$ & $m_B$ & Prob$_{\rm QSO}$(\%) 
& Rad./X flux & peculiarity 
& Observation & Spect.\ ident. & Spect.\ lines & Redsh. \\ \hline

1 & 00 57 19.7 , -48 54 07  & 17.1 & 80 & Rad.: 77+392 & a) PGC 3436/171 &
3600/Oct.8'05/CA  & 
QSO & E: MgII$^b$, H$_\delta ^b$, NeIII & 0.719 \\

2 & 01 00 22.8, -33 41 27 & 17.2 & 66 & Rad.: 3 & a) PGC 3589/180 &
SIMBAD-database &
Star &  & \\

3 & 01 23 46.8, -58 47 44 & 15.7 & 71 & X-ray: 60 & a) PGC 5106/33 
& 3600$^{np}$/Oct.5'05/CA  & 
Star & A: CaII, H$_\alpha $, H$_\beta $, MgI &  \\

4 & 02 26 40.0, -05 52 37 & 14.3 & 99 & Rad.: 107 & d) 
 & 1200$^{np}$/Oct.4'05/CA & 
Star & A: H$_\beta $, MgI & \\

5 & 02 49 58.7, -55 03 20 & 15.4 & 95 & X-ray: 137 & b) NGC 1136/552 
& 600/Oct.7'05/CA  & 
QSO & E: H$_\gamma ^b$, H$_\beta ^b$, OIIIb & 0.230 \\

6 & 02 52 27.8, -20 32 41 & 16.5 & 98 & R: 37, X: 134 & c) PHL 1459/30 
& 3000/Oct.7'05/CA  & 
QSO & E: MgII$^b$, CIII$^b$ & 1.03  \\

7 & 04 25 52.4, -08 33 34 & 16.7 & 71 & X-ray: 3377 & a) PGC 15081/17 
& 5400$^{np}$/Oct.5'05/CA  & 
Seyfert 2 & E: H$_\alpha $, HeII & 0.038   \\

8 & 04 41 47.1, -01 17 56 & 17.1 & 73 & X-ray: 299 & a) PGC 15833/13  
&  10200$^{np}$/Oct.6'05/CA & 
HII-reg. & E: H$_\alpha $, NIIb & 0.029  \\

9 & 04 54 45.4, -18 08 34 & 16.3 & 55 & X-ray: 192 & a) PGC 16315/138 
 &  700/Oct.7'05/CA  & 
Star & A: MgI, NaI &  \\

10 & 05 01 37.6, -04 15 57 & 17.7 & 63 & X-ray: 27 & a) PGC16573/29 
& SIMBAD-database &
Part of a galaxy & &  \\
 
11 & 05 27 16.4, +01 48 40 & 14.8 & 40 & Rad: 27 & d) 
& 900/Nov.1'05/OH &
White dwarf? & A: H$_\alpha ^b$, H$_\beta ^b$, H$_\gamma ^b$, H$_\delta ^b$  &  \\

12 & 05 35 47.6, -05 10 30 & 14.3 & 54 & X-ray: 96 & d) 
& 900/Nov.1'05/OH \& SIMBAD &
Star & E: H$_\alpha $, H$_\beta $, H$_\gamma $, H$_\delta $; A: MgI & \\

13 & 05 35 54.4, -04 48 05 & 15.1 & 75 & X-ray: 105 & d) 
& 900/Nov.1'05/OH \& SIMBAD &
Star  & A: H$_{\alpha ,\beta , \gamma , \delta }$,  
OI, SIIa,b, MgI, NaI & \\ 

14 & 05 36 06.3, -05 41 56 & 14.2 & 54 & X-ray: 88 & d)
& 900/Nov.1'05/OH \& SIMBAD &
Star  & A: H$_\alpha $, H$_\beta $, H$_\delta $, MgI, NaI &   \\

15 & 05 42 32.0, +15 25 05 & 14.8 & 49 & X-ray: 438 & d) 
& 900/Nov.1'05/OH \& SIMBAD &
Cold Star (M5) & E: H$_\alpha $, H$_\beta $, H$_\gamma $, H$_\delta $ 
&  \\

16 & 06 33 50.4, +10 41 23 & 13.8 & 74 & X-ray: 548 & d) 
& 300/Mar.17'05/OH &
White dwarf & A: H$_\alpha $, H$_\beta $, H$_\gamma $, H$_\delta $  &  \\

17 & 07 05 11.8, +37 18 42 & 15.1 & 71 & Rad: 33+65 & d) 
 & 600/Mar.17'05/OH &
Star & A: H$_\alpha $, H$_\beta $, H$_\gamma $ & \\

18 & 08 19 28.0, +70 42 21 & 13.4 & 43 & X-ray: 723 & a)b) PGC 23324/113 
& SIMBAD-database &
HII-reg. & & \\

19 & 08 20 08.2, +34 59 54 & 14.7 & 96 & Rad: 4 & b) PGC 23341/593 
& 600/Mar.15'05/OH \& SIMBAD &
Star & A: H$_\alpha $, H$_\beta $, MgI, NaI &  \\

20 & 08 41 03.8, +03 21 17 & 15.1 & 51 & X-ray: 4345 & d) 
& 600/Mar.15'05/OH & 
White dwarf & A: H$_\alpha ^b$, H$_\beta ^b$, H$_\gamma ^b$, H$_\delta ^b$  
&  \\

21 & 08 43 08.0, +10 44 19 & 17.8 & 97 & Rad: 7 & a) PGC 24492/149 
 & SIMBAD-database &
QSO & & 0.507 \\

22 & 09 28 45.1, +38 31 13 & 18.0 & 52 & X-ray: 187 & a) PGC 26895//110 
& SIMBAD-database &
QSO & & 0.718 \\

23 & 11 47 13.0, +55 42 44 & 17.9 & 66 & X-ray: 293 & a) PGC 36774/142 
& SIMBAD-database &
galaxy & &  \\

24 & 12 04 42.4, +31 10 31 & 17.4 & 83 & X-ray: 92 & a) PGC 38231/68 
& 1800/May 2'06/IU & 
HII-reg. & 
E: H$_{\alpha ,\beta , \gamma , \delta }$, OI,
OII, OIII, HeI, HeII & 0.025 \\     

25 & 12 05 25.8, +51 31 41 & 17.6 & 87 & X-ray: 99 & a) PGC 38295/57 
& SIMBAD-database &
QSO & &  0.504  \\

26 & 12 07 03.4, +65 24 05 & 17.5 & 92 & R: 42, X: 36 & a) PGC 38482/129 
& SIMBAD-database &
QSO & & 1.548 \\ 

27 & 12 21 57.7, +26 04 57 & 17.8 & 95 & X-ray: 228 & a) PGC 40040/162 
 & 1800/May 2'06/IU & 
QSO & E: H$_\beta ^b$, H$_\gamma ^b$, 
OII, OIIIa,b, NeIII  & 0.277 \\

28 & 13 20 14.2, +33 07 53 & 15.7 & 91 & X-ray: 237 & a) PGC 46515/43 
& 1200/May 2'06/IU & 
Star & A: H$_\alpha $, H$_\beta $, H$_\gamma $, H$_\delta $  & \\

29 & 13 42 10.6, +60 47 40 & 17.9 & 97 & Rad: 2 & a) PGC 48534/75 
& SIMBAD-database &
QSO & & 1.225 \\

30 & 14 17 58.8, +25 05 33 & 15.9 & 69 & X-ray: 5645 & a) PGC 51074/160 
& SIMBAD-database &
Star & &  \\

%
%
%
%

31 & 14 26 18.6 , +26 14 52 & 17.2 & 77 & X-ray: 59 & a) PGC 51549/48 
& 1800/May 2'06/IU & 
HII-reg. & E: H$_\alpha $, NIIa,b  & 0.030 \\

32 & 14 40 06.0, +75 05 33 & 14.8 & 88 & X-ray: 1655 & d) 
& 600/Mar.15'05/OH \& SIMBAD & 
White dwarf & A: H$_\alpha ^b$, H$_\beta ^b$, H$_\gamma ^b$, H$_\delta ^b$  
&  \\

33 & 16 12 50.8, +00 02 53 & 15.2 & 73 & Rad: 39+54 & b) PGC 57505/531 
& 900/Jul.17'06/CA & 
Star & A: H$_\alpha $, H$_\beta $ & \\

34 & 17 37 42.6, -59 53 41 & 17.3 & 42 & X-ray: 7 & a) PGC 60594/170 
& 5400/Jul.16,17'06/CA & Star & A: H$_\beta $ & \\

35 & 18 24 36.2, +73 10 45 & 17.5 & 94 & X-ray: 7 & a) PGC 61833/120 
& 1800/May 2'06/IU & 
QSO & A: CIV$^b$, CIII$^b$  & 1.96 \\

36 & 20 09 38.2, -49 17 45 & 17.0 & 86 & X-ray: 47  & a) PGC 64185/126 
&  2000/Oct.8'05/CA  & 
QSO & E: MgII$^b$, CIII$^b$ & 0.998 \\

37 & 20 12 42.6, -54 00 29 & 14.4 & 64 & X-ray: 156 & b) PGC 64288/332 
& 1800/Jul.16'06/CA & 
Star & A: H$_\alpha $, H$_\beta $ &  \\

38 & 20 43 46.6, -26 35 03 & 17.8 & 74 & Rad.: 20+15 & a) PGC 65268/121 
& 3000/Oct.8'05/CA  & 
Star & A: H$_\alpha $, MgI &  \\

39 & 22 37 49.3, +23 45 41 & 17.7 & 46 & X-ray: 10 & a) PGC 69364/99 
& 2000/Sep.25'06/OH & QSO & E: MgII$^b$, CIII$^b$ & 1.24 \\

40 & 22 57 52.3, +26 06 38 & 17.9 & 67 & Rad: 19 & a) PGC 70116/100 
& 4000/Sep.25,26'06/OH & Star & A: MgI, NaI & \\

41 & 23 38 41.6, +26 48 04 & 16.9 & 76 & X-ray: 35 & a) PGC 71995/124 
 & SIMBAD-database & Star & & \\ 
\end{tabular}
\end{center}
\end{table}

\end{landscape}

\end{document}